# Four-dimensional equation of motion for viscous compressible and charged fluid with regard to the acceleration field, pressure field and dissipation field

Sergey G. Fedosin

PO box 614088, Sviazeva str. 22-79, Perm, Russia

E-mail: intelli@list.ru

From the principle of least action the equation of motion for viscous compressible and charged fluid is derived. The viscosity effect is described by the 4-potential of the energy dissipation field, dissipation tensor and dissipation stress-energy tensor. In the weak field limit it is shown that the obtained equation is equivalent to the Navier-Stokes equation. The equation for the power of the kinetic energy loss is provided, the equation of motion is integrated, and the dependence of the velocity magnitude is determined. A complete set of equations is presented, which suffices to solve the problem of motion of viscous compressible and charged fluid in the gravitational and electromagnetic fields.

## 1. Introduction

Since Navier-Stokes equations appeared in 1827 [1], [2], constant attempts have been made to derive these equations by various methods. Stokes [3] and Saint-Venant [4] in derivation of these equations relied on the fact that the deviatoric stress tensor of normal and tangential stress is linearly related to the three-dimensional deformation rate tensor and the fluid is isotropic.

In book [5] it is considered that Navier-Stokes equations are the extremum conditions of some functional, and a method of finding a solution of these equations is described, which consists in the gradient motion to the extremum of this functional.

One of the variants of the four-dimensional stress-energy tensor of viscous stresses in the special theory of relativity can be found in [6]. The divergence of this tensor gives the required viscous terms in the Navier-Stokes equation. The phenomenological derivation of this tensor is based on the assumed condition of entropy increment during energy dissipation. As a consequence, in the co-moving reference frame the time components of the tensor, i.e. the dissipation energy density and its flux vanish. Therefore, such a tensor is not a universal tensor and cannot serve, for example, as a basis for determining the metric in the presence of viscosity.

In this article, our goal is to provide in general form the four-dimensional stress-energy tensor of energy dissipation, which describes in the curved spacetime the energy density and the stress and energy flux, arising due to viscous stresses. This tensor will be derived with the help of the principle of least action on the basis of a covariant 4-potential of the dissipation field. Then we will apply these quantities in the equation of motion of the viscous compressible and charged fluid, and by selecting the scalar potential of the dissipation field we will obtain the Navier-Stokes equation. The essential element of our calculations will be the use of the wave equation for the field potentials of the acceleration field. In conclusion, we will present a complete set of equations sufficient to describe the motion of viscous fluid.

The summary of notation for all the fields is provided in the Table.

|  | Electromagnetic field | Gravitational field | Acceleration field | Pressure field | Dissipation field |
|---|---|---|---|---|---|
| 4-potential | $A_\mu$ | $D_\mu$ | $u_\mu$ | $\pi_\mu$ | $\lambda_\mu$ |
| Scalar potential | $\varphi$ | $\psi$ | $\vartheta$ | $\wp$ | $\varepsilon$ |
| Vector potential | **A** | **D** | **U** | **Π** | **Θ** |
| Field strength | **E** | **Γ** | **S** | **C** | **X** |
| Solenoidal vector | **B** | **Ω** | **N** | **I** | **Y** |
| Field tensor | $F_{\mu\nu}$ | $\Phi_{\mu\nu}$ | $u_{\mu\nu}$ | $f_{\mu\nu}$ | $h_{\mu\nu}$ |
| Stress-energy tensor | $W^{\alpha\beta}$ | $U^{\alpha\beta}$ | $B^{\alpha\beta}$ | $P^{\alpha\beta}$ | $Q^{\alpha\beta}$ |
| Energy-momentum flux vector | **P** | **H** | **K** | **F** | **Z** |

In this Table **P** is the Poynting vector, **H** is the Heaviside vector. For simplicity, we will assume that the various fields existing simultaneously would not produce any induced effects (currents) due to coupling and interactions between the fields are absent.

## 2. The action function

The starting point of our calculations is the action function in the following form:

$$S = \int L\,dt = \int \left( \begin{array}{l} k(R-2\Lambda) - \dfrac{1}{c} D_\mu J^\mu + \dfrac{c}{16\pi G}\Phi_{\mu\nu}\Phi^{\mu\nu} - \dfrac{1}{c}A_\mu j^\mu - \dfrac{c\varepsilon_0}{4}F_{\mu\nu}F^{\mu\nu} - \dfrac{1}{c}u_\mu J^\mu \\ -\dfrac{c}{16\pi\eta}u_{\mu\nu}u^{\mu\nu} - \dfrac{1}{c}\pi_\mu J^\mu - \dfrac{c}{16\pi\sigma}f_{\mu\nu}f^{\mu\nu} - \dfrac{1}{c}\lambda_\mu J^\mu - \dfrac{c}{16\pi\tau}h_{\mu\nu}h^{\mu\nu} \end{array} \right) \sqrt{-g}\,d\Sigma,$$

(1)

where $L$ is the Lagrange function or Lagrangian,

$R$ is the scalar curvature,

$\Lambda$ is the cosmological constant,

$J^\mu = \rho_0 u^\mu$ is the 4-vector of the mass (gravitational) current,

$\rho_0$ is the mass density in the reference frame associated with the fluid unit,

$u^\mu = \dfrac{c\,dx^\mu}{ds}$ is the 4-velocity of a point particle, $c$ is the speed of light,

$D_\mu = \left(\dfrac{\psi}{c}, -\mathbf{D}\right)$ is the 4-potential of the gravitational field, described by the scalar potential $\psi$ and the vector potential **D** of this field,

$G$ is the gravitational constant,

$\Phi_{\mu\nu}$ is the gravitational tensor,

$A_\mu = \left(\dfrac{\varphi}{c}, -\mathbf{A}\right)$ is the 4-potential of the electromagnetic field, which is specified by the scalar potential $\varphi$ and the vector potential **A** of this field,

$j^\mu = \rho_{0q} u^\mu$ is the 4-vector of the electromagnetic (charge) current,

$\rho_{0q}$ is the charge density in the reference frame associated with the fluid unit,

$\varepsilon_0$ is the vacuum permittivity,

$F_{\mu\nu}$ is the electromagnetic tensor,

$u_\mu = g_{\mu\nu} u^\nu$ is the 4-velocity with the covariant index, expressed with the help of the metric tensor and the 4-velocity with the contravariant index; the covariant 4-velocity is the 4-potential of the acceleration field $u_\mu = \left( \dfrac{\vartheta}{c}, -\mathbf{U} \right)$, where $\vartheta$ and $\mathbf{U}$ denote the scalar and vector potentials, respectively,

$u_{\mu\nu}$ is the acceleration tensor,

$\eta$, $\sigma$ and $\tau$ are some functions of coordinates and time,

$\pi_\mu = \dfrac{p_0}{\rho_0 c^2} u_\mu = \left( \dfrac{\wp}{c}, -\mathbf{\Pi} \right)$ is the 4-potential of the pressure field, consisting of the scalar potential $\wp$ and the vector potential $\mathbf{\Pi}$, $p_0$ is the pressure in the reference frame associated with the particle, the ratio $\dfrac{p_0}{\rho_0 c^2}$ defines the equation of state of the fluid,

$f_{\mu\nu}$ is the pressure field tensor.

The above-mentioned quantities are described in detail in [7]. In addition to them, we introduce the 4-potential of energy dissipation in the fluid:

$$\lambda_\mu = \dfrac{\alpha\, u_\mu}{c^2} = \left( \dfrac{\varepsilon}{c}, -\mathbf{\Theta} \right), \qquad (2)$$

where $\alpha$ is the dissipation function, $\varepsilon$ and $\mathbf{\Theta}$ are the scalar and vector dissipation potentials, respectively.

Using the 4-potential $\lambda_\mu$ we construct the energy dissipation tensor:

$$h_{\mu\nu} = \nabla_\mu \lambda_\nu - \nabla_\nu \lambda_\mu = \partial_\mu \lambda_\nu - \partial_\nu \lambda_\mu. \qquad (3)$$

The coefficients $\eta$, $\sigma$ and $\tau$ in order to simplify calculations we assume to be a constants.

The term $-\frac{1}{c}\lambda_\mu J^\mu$ in (1) reflects the fact that the energy of the fluid motion can be dissipated in the surrounding medium and turn into the internal fluid energy, while the system's energy does not change. The last term in (1) is associated with the energy, accumulated by the system due to the action of the energy dissipation.

The method of constructing the dissipation 4-potential $\lambda_\mu$ in (2) and the dissipation tensor $h_{\mu\nu}$ in (3) is fully identical to that, which was used earlier in [7]. Therefore, we will not provide here the intermediate results from [7], and will right away write the equations of motion of the fluid and field, obtained as a result of the variation of the action function (1).

### 3. Field equations

The electromagnetic field equations have the standard form:

$$\nabla_\sigma F_{\mu\nu} + \nabla_\nu F_{\sigma\mu} + \nabla_\mu F_{\nu\sigma} = 0, \qquad \nabla_\beta F^{\alpha\beta} = -\mu_0 j^\alpha, \tag{4}$$

where $\mu_0 = \dfrac{1}{c^2 \varepsilon_0}$ is the vacuum permeability.

The gravitational field equations are:

$$\nabla_\sigma \Phi_{\mu\nu} + \nabla_\nu \Phi_{\sigma\mu} + \nabla_\mu \Phi_{\nu\sigma} = 0, \qquad \nabla_\beta \Phi^{\alpha\beta} = \frac{4\pi G}{c^2} J^\alpha. \tag{5}$$

The acceleration field equations are:

$$\nabla_\sigma u_{\mu\nu} + \nabla_\nu u_{\sigma\mu} + \nabla_\mu u_{\nu\sigma} = 0, \qquad \nabla_\beta u^{\alpha\beta} = -\frac{4\pi\eta}{c^2} J^\alpha. \tag{6}$$

The pressure field equations are:

$$\nabla_\sigma f_{\mu\nu} + \nabla_\nu f_{\sigma\mu} + \nabla_\mu f_{\nu\sigma} = 0, \qquad \nabla_\beta f^{\alpha\beta} = -\frac{4\pi\sigma}{c^2} J^\alpha. \tag{7}$$

The dissipation field equations are:

$$\nabla_\sigma h_{\mu\nu} + \nabla_\nu h_{\sigma\mu} + \nabla_\mu h_{\nu\sigma} = 0, \qquad \nabla_\beta h^{\alpha\beta} = -\frac{4\pi\tau}{c^2} J^\alpha. \qquad (8)$$

In order to obtain the equations (4-8), variation by the corresponding 4-potential is carried out in the action function (1).

All the above-mentioned fields have vector character. Each field can be described by two three-dimensional vectors, included into the corresponding field tensor. One of these vectors is the strength of the corresponding field, and the other solenoidal vector describes the field vorticity. For example, the components of the electric field strength **E** and the magnetic field induction **B** are the components of the electromagnetic tensor $F_{\mu\nu}$. The gravitational tensor $\Phi_{\mu\nu}$ consists of the components of the gravitational field strength **Γ** and the torsion field **Ω**.

As can be seen from (4-8), the constants $\eta$, $\sigma$ and $\tau$ have the same meaning as the constants $\mu_0$ and $G$ – all these constants reflect the relationship between the 4-current of dissipated fluid and the divergence of the corresponding field tensor. The properties of the dissipation field are provided in Appendix A.

### 4. Field gauge

In order to simplify the form of equations we use the following field gauge:

$$\nabla_\beta A^\beta = \nabla^\mu A_\mu = 0, \qquad \nabla_\beta D^\beta = \nabla^\mu D_\mu = 0, \qquad (9)$$

$$\nabla_\beta u^\beta = \nabla^\mu u_\mu = 0, \qquad \nabla_\beta \pi^\beta = \nabla^\mu \pi_\mu = 0, \qquad \nabla_\beta \lambda^\beta = \nabla^\mu \lambda_\mu = 0.$$

In (9) the gauge of each field is carried out by equating the covariant derivative of the corresponding 4-potential to zero. Since the 4-potentials consist of the scalar and vector potentials, the gauge (9) links the scalar and vector potential of each field. As a result, the divergence of the vector potential of any field in a certain volume is accompanied by change in time of the scalar field potential in this volume, and also depends on the tensor product of the Christoffel symbols and the 4-potential, that is, on the degree of spacetime curvature.

## 5. Continuity equations

In equations (4-8) the divergences of field tensors are associated with their sources, i.e. with 4-currents. The field tensors are defined by their 4-potentials similarly to (3):

$$F_{\mu\nu} = \nabla_\mu A_\nu - \nabla_\nu A_\mu = \partial_\mu A_\nu - \partial_\nu A_\mu, \qquad \Phi_{\mu\nu} = \nabla_\mu D_\nu - \nabla_\nu D_\mu = \partial_\mu D_\nu - \partial_\nu D_\mu, \qquad (10)$$

$$u_{\mu\nu} = \nabla_\mu u_\nu - \nabla_\nu u_\mu = \partial_\mu u_\nu - \partial_\nu u_\mu, \qquad f_{\mu\nu} = \nabla_\mu \pi_\nu - \nabla_\nu \pi_\mu = \partial_\mu \pi_\nu - \partial_\nu \pi_\mu.$$

If we substitute (3) and (10) into equations (4-8), and apply the covariant derivative $\nabla_\alpha$ to all terms, we obtain the following relations containing the Ricci tensor:

$$R_{\mu\alpha} F^{\mu\alpha} = \mu_0 \nabla_\alpha j^\alpha, \quad R_{\mu\alpha} \Phi^{\mu\alpha} = -\frac{4\pi G}{c^2} \nabla_\alpha J^\alpha, \quad R_{\mu\alpha} u^{\mu\alpha} = \frac{4\pi \eta}{c^2} \nabla_\alpha J^\alpha,$$

$$R_{\mu\alpha} f^{\mu\alpha} = \frac{4\pi \sigma}{c^2} \nabla_\alpha J^\alpha, \qquad R_{\mu\alpha} h^{\mu\alpha} = \frac{4\pi \tau}{c^2} \nabla_\alpha J^\alpha. \qquad (11)$$

In the limit of special theory of relativity, the Ricci tensor vanishes, the covariant derivative turns into the 4-gradient, and then instead of (11) we can write:

$$\partial_\alpha j^\alpha = 0, \qquad \partial_\alpha J^\alpha = 0. \qquad (12)$$

Relations (12) are the ordinary continuity equation of the charge and mass 4-currents in the flat spacetime.

## 6. Equations of motion

The variation of the action function leads directly to the equations describing the motion of the fluid unit under the action of the fields:

$$-u_{\beta\sigma} J^\sigma = \Phi_{\beta\sigma} J^\sigma + F_{\beta\sigma} j^\sigma + f_{\beta\sigma} J^\sigma + h_{\beta\sigma} J^\sigma. \qquad (13)$$

The left side of the equality can be transformed, taking into account the expression $J^\sigma = \rho_0 u^\sigma$ for the 4-vector of the mass current density and the definition (10) for the acceleration tensor $u_{\beta\sigma} = \nabla_\beta u_\sigma - \nabla_\sigma u_\beta$:

$$-u_{\beta\sigma} J^\sigma = -\rho_0 u^\sigma \left( \nabla_\beta u_\sigma - \nabla_\sigma u_\beta \right) = \rho_0 u^\sigma \nabla_\sigma u_\beta = \rho_0 \frac{D u_\beta}{D\tau} = \rho_0 \frac{d u_\beta}{d\tau} - \rho_0 \Gamma^\lambda_{\sigma\beta} u_\lambda u^\sigma = \rho_0 a_\beta.$$

(14)

In (14) $a_\beta$ denotes the 4-acceleration, and we used the operator of proper-time-derivative $u^\sigma \nabla_\sigma = \frac{D}{D\tau}$, where $D$ is the symbol of 4-differential in the curved spacetime, $\tau$ is the proper time [8]. If we substitute (14) into (13), we obtain the equation of motion, in which the 4-acceleration is expressed in terms of field tensors and 4-currents:

$$\rho_0 a_\beta = \Phi_{\beta\sigma} J^\sigma + F_{\beta\sigma} j^\sigma + f_{\beta\sigma} J^\sigma + h_{\beta\sigma} J^\sigma.$$

(15)

Variation of the action function allows us to find the form of stress-energy tensors of all the fields associated with the fluid:

$$W^{\alpha\beta} = \varepsilon_0 c^2 \left( -g^{\alpha\nu} F_{\kappa\nu} F^{\kappa\beta} + \frac{1}{4} g^{\alpha\beta} F_{\mu\nu} F^{\mu\nu} \right), \quad U^{\alpha\beta} = -\frac{c^2}{4\pi G} \left( -g^{\alpha\nu} \Phi_{\kappa\nu} \Phi^{\kappa\beta} + \frac{1}{4} g^{\alpha\beta} \Phi_{\mu\nu} \Phi^{\mu\nu} \right),$$

$$B^{\alpha\beta} = \frac{c^2}{4\pi\eta} \left( -g^{\alpha\nu} u_{\kappa\nu} u^{\kappa\beta} + \frac{1}{4} g^{\alpha\beta} u_{\mu\nu} u^{\mu\nu} \right), \quad P^{\alpha\beta} = \frac{c^2}{4\pi\sigma} \left( -g^{\alpha\nu} f_{\kappa\nu} f^{\kappa\beta} + \frac{1}{4} g^{\alpha\beta} f_{\mu\nu} f^{\mu\nu} \right),$$

$$Q^{\alpha\beta} = \frac{c^2}{4\pi\tau} \left( -g^{\alpha\nu} h_{\kappa\nu} h^{\kappa\beta} + \frac{1}{4} g^{\alpha\beta} h_{\mu\nu} h^{\mu\nu} \right).$$

(16)

One of the properties of these tensors is that their divergences alongside with the field tensors specify the densities of 4-forces, arising from the influence of the corresponding field on the fluid:

$$(f_\beta)_e = F_{\beta\sigma} j^\sigma = -\nabla^k W_{\beta k}, \qquad (f_\beta)_g = \Phi_{\beta\sigma} J^\sigma = -\nabla^k U_{\beta k},$$

$$(f_\beta)_a = \rho_0 a_\beta = -u_{\beta\sigma} J^\sigma = \nabla^k B_{\beta k}, \qquad (f_\beta)_p = f_{\beta\sigma} J^\sigma = -\nabla^k P_{\beta k},$$

$$(f_\beta)_d = h_{\beta\sigma} J^\sigma = -\nabla^k Q_{\beta k}. \tag{17}$$

The left side of (17) contains the density of the corresponding 4-force, excluding $(f_\beta)_a$, which up to sign denotes the density of the 4-force, acting from the accelerated fluid on the rest four fields.

From (13-17) it follows that the equation of motion can be written only in terms of the divergences of the stress-energy tensors of fields:

$$\nabla_\beta (W^{\alpha\beta} + U^{\alpha\beta} + B^{\alpha\beta} + P^{\alpha\beta} + Q^{\alpha\beta}) = 0. \tag{18}$$

We have integrated equation (18) in [9] in the weak field limit (excluding the stress-energy tensor of dissipation $Q^{\alpha\beta}$), and this allowed us to explain the well-known 4/3 problem of the fields mass-energy inequality in the fixed and moving systems. The equation (18) will be used also in the equation (21) for the metric.

### 7. The system's energy

The action function (1) contains the Lagrangian $L$. Applying to it the Legendre transformations for a system of fluid units, we can find the system's Hamiltonian. This Hamiltonian is the relativistic energy of the system, written in an arbitrary reference frame. Since the energy is dependent on the cosmological constant $\Lambda$, gauging of the cosmological constant should be done using the relation:

$$ck\Lambda = -D_\mu J^\mu - A_\mu j^\mu - u_\mu J^\mu - \pi_\mu J^\mu - \lambda_\mu J^\mu. \tag{19}$$

As a result, we can write for the energy of the system the following:

$$E = \frac{1}{c}\int \left(\rho_{0q}\varphi + \rho_0\psi + \rho_0\vartheta + \rho_0\wp + \rho_0\varepsilon\right) u^0 \sqrt{-g}\, dx^1 dx^2 dx^3 -$$

$$-\int \left(-\frac{1}{4\mu_0}F_{\mu\nu}F^{\mu\nu} + \frac{c^2}{16\pi G}\Phi_{\mu\nu}\Phi^{\mu\nu} - \frac{c^2}{16\pi\eta}u_{\mu\nu}u^{\mu\nu} - \frac{c^2}{16\pi\sigma}f_{\mu\nu}f^{\mu\nu} - \frac{c^2}{16\pi\tau}h_{\mu\nu}h^{\mu\nu}\right)\sqrt{-g}\, dx^1 dx^2 dx^3.$$

The energy of the system in the form of a set of closely interacting particles and the related fields in the weak field limit was calculated in [10]. The difference between the system's mass and the gravitational mass was shown, as well as the fact that the mass-energy of the proper electromagnetic field reduces the gravitational mass of the system.

### 8. Equation for the metric

According to the logic of the covariant theory of gravitation [11] and the metric theory of relativity [12], contribution to the definition of the system's metric is made by the stress-energy tensors of all the fields, including the gravitational field. The metric is a secondary function, the derivative of the fields acting in the system that define all the basic properties of the system. The equation for the metric is obtained as follows:

$$R^{\alpha\beta} - \frac{1}{4} R g^{\alpha\beta} = -\frac{1}{2ck}(W^{\alpha\beta} + U^{\alpha\beta} + B^{\alpha\beta} + P^{\alpha\beta} + Q^{\alpha\beta}). \qquad (21)$$

If we multiply (21) by the metric tensor $g_{\alpha\beta}$ and contract over all the indices, the right and left sides of the equation vanish. It follows from the properties of tensors in (21). Outside the fluid limits, with regard to the gauge, the scalar curvature $R$ becomes equal to zero. If we take into account the equation of motion (18), the covariant derivative of the right side of (21) is zero. The covariant derivative of the left side of (21) is also zero, since $\nabla_\beta R = 0$ as a consequence of the cosmological constant gauge, and for the Einstein tensor the following equality holds: $\nabla_\beta \left( R^{\alpha\beta} - \frac{1}{2} R g^{\alpha\beta} \right) = 0$.

### 9. The analysis of the equation of motion

Equations (14-15) imply the connection between the covariant 4-acceleration of a fluid unit and the densities of acting forces in the curved spacetime:

$$\rho_0 a_\beta = \rho_0 \frac{d u_\beta}{d\tau} - \rho_0 \Gamma^\lambda_{\sigma\beta} u_\lambda u^\sigma = \Phi_{\beta\sigma} J^\sigma + F_{\beta\sigma} j^\sigma + f_{\beta\sigma} J^\sigma + h_{\beta\sigma} J^\sigma.$$

We will write this four-dimensional equation separately for the time and space components, given that

$$J^\sigma = c\rho_0 \frac{dx^\sigma}{ds} = c\rho_0 \frac{dt}{ds}\frac{dx^\sigma}{dt} = c\rho_0 \frac{dt}{ds}(c, \mathbf{v}),$$

as well as $j^\sigma = c\rho_{0q}\frac{dx^\sigma}{ds} = c\rho_{0q}\frac{dt}{ds}(c, \mathbf{v})$. For the dissipation field we will use relations (A6) from Appendix A, where in the general case $\frac{cdt}{ds}$ should be substituted instead of $\gamma$. Expressions for other fields can be found in [7]. This gives:

$$c\rho_0 \frac{du_0}{ds} - \rho_0 \Gamma^\lambda_{\sigma 0} u_\lambda u^\sigma = \rho_0 \frac{dt}{ds}\left(\mathbf{\Gamma}\cdot\mathbf{v} + \frac{\rho_{0q}}{\rho_0}\mathbf{E}\cdot\mathbf{v} + \mathbf{C}\cdot\mathbf{v} + \mathbf{X}\cdot\mathbf{v}\right).$$

$$c\rho_0 \frac{du_i}{ds} - \rho_0 \Gamma^\lambda_{\sigma i} u_\lambda u^\sigma = -c\rho_0 \frac{dt}{ds}\left(\mathbf{\Gamma} + [\mathbf{v}\times\mathbf{\Omega}] + \frac{\rho_{0q}}{\rho_0}\left(\mathbf{E} + [\mathbf{v}\times\mathbf{B}]\right) + \mathbf{C} + [\mathbf{v}\times\mathbf{I}] + \mathbf{X} + [\mathbf{v}\times\mathbf{Y}]\right).$$

Here $\mathbf{\Gamma}$, $\mathbf{E}$, $\mathbf{C}$ and $\mathbf{X}$ are the vectors of strengths of gravitational and electromagnetic fields, pressure field and dissipation field, respectively. Notations $\mathbf{\Omega}$, $\mathbf{B}$, $\mathbf{I}$ and $\mathbf{Y}$ refer to the torsion field, the magnetic field and the solenoidal vectors of the pressure and dissipation fields, respectively.

After reduction by a factor $\rho_0 \frac{dt}{ds}$ we obtain:

$$c\frac{du_0}{dt} - \frac{ds}{dt}\Gamma^\lambda_{\sigma 0}u_\lambda u^\sigma = \mathbf{\Gamma}\cdot\mathbf{v} + \frac{\rho_{0q}}{\rho_0}\mathbf{E}\cdot\mathbf{v} + \mathbf{C}\cdot\mathbf{v} + \mathbf{X}\cdot\mathbf{v}. \qquad (22)$$

$$\frac{du_i}{dt} - \frac{d\tau}{dt}\Gamma^\lambda_{\sigma i}u_\lambda u^\sigma = -\mathbf{\Gamma} - [\mathbf{v}\times\mathbf{\Omega}] - \frac{\rho_{0q}}{\rho_0}\mathbf{E} - \frac{\rho_{0q}}{\rho_0}[\mathbf{v}\times\mathbf{B}] - \mathbf{C} - [\mathbf{v}\times\mathbf{I}] - \mathbf{X} - [\mathbf{v}\times\mathbf{Y}]. \qquad (23)$$

In (23) the sum $\frac{\rho_{0q}}{\rho_0}\mathbf{E} + \frac{\rho_{0q}}{\rho_0}[\mathbf{v}\times\mathbf{B}]$ is the contribution of the electromagnetic Lorentz force, acting on the fluid unit, into the total acceleration. The minus sign before this sum appears because $u_i$ is the space component of the covariant 4-velocity, which differs from the ordinary contravariant space component $u^i$ in the form factor of the metric tensor. Similarly, the sum $\mathbf{\Gamma} + [\mathbf{v}\times\mathbf{\Omega}]$ is the acceleration of the gravitational Lorentz force.

Gravitational and electromagnetic forces are the so-called mass forces distributed over the entire volume, where there is mass and charge of the fluid.

### 9.1. The equation of motion in Minkowski space

In order to simplify our analysis, we will consider equations (22-23) in the framework of the special theory of relativity. The sum of the last two terms in (23), taking into account the formulas (A7) from Appendix A, for $\mathbf{X}$ and $\mathbf{Y}$ gives the following:

$$\mathbf{X}+[\mathbf{v}\times\mathbf{Y}]=-\nabla(\alpha\gamma)-\frac{1}{c^2}\frac{\partial(\alpha\gamma\mathbf{v})}{\partial t}+\frac{1}{c^2}\mathbf{v}\times[\nabla\times(\alpha\gamma\mathbf{v})]=$$

$$=-\frac{1}{2\alpha\gamma c^2}\nabla(\alpha\gamma c)^2-\frac{1}{c^2}\frac{\partial(\alpha\gamma\mathbf{v})}{\partial t}-\frac{1}{c^2}(\mathbf{v}\cdot\nabla)(\alpha\gamma\mathbf{v})+\frac{1}{2\alpha\gamma c^2}\nabla(\alpha\gamma\mathbf{v})^2=-\frac{1}{\gamma}\nabla\alpha-\frac{1}{c^2}\frac{d(\alpha\gamma\mathbf{v})}{dt}.$$

(24)

In (24) $\gamma=\dfrac{1}{\sqrt{1-v^2/c^2}}=\dfrac{cdt}{ds}$ is the Lorentz factor.

For the pressure field, with regard to the definition of the 4-potential in the form of $\pi_\mu=\dfrac{p_0}{\rho_0 c^2}u_\mu=\left(\dfrac{\wp}{c},-\mathbf{\Pi}\right)$, the pressure field tensor $f_{\mu\nu}$ from (10) and the definition of the vectors $\mathbf{C}$ and $\mathbf{I}$ by the rule:

$$f_{0i}=\partial_0\pi_i-\partial_i\pi_0=\frac{1}{c}C_i, \qquad f_{ij}=\partial_i\pi_j-\partial_j\pi_i=-I_k,$$

we find the expression for the vectors:

$$\mathbf{C}=-\nabla\wp-\frac{\partial\mathbf{\Pi}}{\partial t}=-\nabla\left(\frac{\gamma p_0}{\rho_0}\right)-\frac{\partial}{\partial t}\left(\frac{\gamma p_0 \mathbf{v}}{\rho_0 c^2}\right), \qquad \mathbf{I}=\nabla\times\mathbf{\Pi}=\nabla\times\left(\frac{\gamma p_0 \mathbf{v}}{\rho_0 c^2}\right). \qquad (25)$$

Using (25) we calculate the sum of two terms in (23):

$$\mathbf{C}+[\mathbf{v}\times \mathbf{I}] = -\nabla\left(\frac{\gamma p_0}{\rho_0}\right) - \frac{\partial}{\partial t}\left(\frac{\gamma p_0 \mathbf{v}}{\rho_0 c^2}\right) + \mathbf{v}\times\left[\nabla\times\left(\frac{\gamma p_0 \mathbf{v}}{\rho_0 c^2}\right)\right] =$$

$$= -\frac{\rho_0}{2\gamma p_0}\nabla\left(\frac{\gamma p_0}{\rho_0}\right)^2 - \frac{\partial}{\partial t}\left(\frac{\gamma p_0 \mathbf{v}}{\rho_0 c^2}\right) - (\mathbf{v}\cdot\nabla)\left(\frac{\gamma p_0 \mathbf{v}}{\rho_0 c^2}\right) + \frac{\rho_0}{2\gamma p_0}\nabla\left(\frac{\gamma p_0 \mathbf{v}}{\rho_0 c}\right)^2 = -\frac{1}{\gamma}\nabla\left(\frac{p_0}{\rho_0}\right) - \frac{d}{dt}\left(\frac{\gamma p_0 \mathbf{v}}{\rho_0 c^2}\right).$$

(26)

Substituting (24) and (26) into (23), and taking into account that in Minkowski space the Christoffel symbols are zero and the space component of the 4-velocity equals $u_i = -\gamma \mathbf{v}$, we find:

$$\frac{d}{dt}\left[\gamma \mathbf{v}\left(1 + \frac{p_0}{\rho_0 c^2} + \frac{\alpha}{c^2}\right)\right] = \mathbf{a}_m - \frac{1}{\gamma}\nabla\left(\frac{p_0}{\rho_0}\right) - \frac{1}{\gamma}\nabla\alpha. \qquad (27)$$

In (27) we have introduced notation for the acceleration, resulting from the action of mass forces:

$$\mathbf{a}_m = \mathbf{\Gamma} + [\mathbf{v}\times\mathbf{\Omega}] + \frac{\rho_{0q}}{\rho_0}\mathbf{E} + \frac{\rho_{0q}}{\rho_0}[\mathbf{v}\times\mathbf{B}].$$

Until now we have not defined the dissipation function $\alpha$. In this approximation, it is associated with a scalar potential $\varepsilon$ of the dissipation field by relation: $\varepsilon = \alpha\gamma$.

Let us assume that

$$\nabla\alpha = \nabla\left(\frac{\varepsilon}{\gamma}\right) = \frac{\varsigma}{\rho_0 c^2}\mathbf{v} - \frac{\omega}{\rho_0}\nabla(\nabla\cdot\mathbf{v}), \qquad (28)$$

that is $\varepsilon = \frac{\gamma\varsigma}{c^2}\int\frac{1}{\rho_0}\mathbf{v}\,d\mathbf{r} - \gamma\omega\int\frac{1}{\rho_0}\nabla(\nabla\cdot\mathbf{v})\,d\mathbf{r}$.

It means that the scalar potential of the dissipation field is proportional both to the velocity $\mathbf{v}$ of the considered fluid unit and the path traveled by it in the surrounding space.

Contribution to $\varepsilon$ is also made by the gradient of the velocity divergence with a certain coefficient $\omega$.

The coefficient $\varsigma$ depends on the parameters of interacting fluid layers, in a first approximation it is inversely proportional to the square of the layers' thickness. At the same time the coefficient $\varsigma$ reflects the fluid properties and can be different in different fluids. Taking into account (28), equation (27) is transformed as follows:

$$\frac{d}{dt}\left[\gamma \mathbf{v}\left(1+\frac{p_0}{\rho_0 c^2}+\frac{\alpha}{c^2}\right)\right] = \mathbf{a}_m - \frac{1}{\gamma}\nabla\left(\frac{p_0}{\rho_0}\right) - \frac{\varsigma}{\gamma\rho_0 c^2}\mathbf{v} + \frac{\omega}{\gamma\rho_0}\nabla(\nabla\cdot\mathbf{v}). \qquad (29)$$

Due to the presence in (29) of the gradient $\nabla\left(\dfrac{p_0}{\rho_0}\right)$ of the pressure to the mass density ratio, there is acceleration directed against this gradient. The term in (29) which is proportional to the velocity $\mathbf{v}$, defines the rate of deceleration due to viscosity. Since the deceleration in (29) depends not on the absolute velocity but on the velocity of motion of some fluid layers relative to the other layers, the velocity $\mathbf{v}$ should be a relative velocity. We will use the freedom of choosing the reference frame in order to move from absolute velocities to relative velocities. Suppose the reference frame is co-moving and it moves in the fluid with the control volume of a small size. Then in such a reference frame the velocity $\mathbf{v}$ in (29) will be a relative velocity: some layers will be ahead, while others will lag behind, and viscous forces will appear.

We will now write the equations for the acceleration field from [7]:

$$\nabla\cdot\mathbf{S} = 4\pi\eta\gamma\rho_0, \quad \nabla\times\mathbf{N} = \frac{1}{c^2}\frac{\partial \mathbf{S}}{\partial t}+\frac{4\pi\eta\gamma\rho_0 \mathbf{v}}{c^2}, \quad \nabla\cdot\mathbf{N}=0, \quad \nabla\times\mathbf{S}=-\frac{\partial \mathbf{N}}{\partial t}. \qquad (30)$$

The vector $\mathbf{S}$ in (30) is the acceleration field strength, and the vector $\mathbf{N}$ is the solenoidal vector of the acceleration field. The 4-potential of the acceleration field $u_\mu = \left(\dfrac{\vartheta}{c}, -\mathbf{U}\right)$ equals the 4-velocity, taken with the covariant index. The acceleration tensor $u_{\mu\nu}$ is defined in (10) as a 4-curl and it contains the vectors $\mathbf{S}$ and $\mathbf{N}$:

$$u_{0i} = \partial_0 u_i - \partial_i u_0 = \frac{1}{c} S_i, \qquad u_{ij} = \partial_i u_j - \partial_j u_i = -N_k.$$

In Minkowski space we can move from the scalar $\vartheta$ and vector $\mathbf{U}$ potentials of the acceleration field to the 4-velocity components and express vectors $\mathbf{S}$ and $\mathbf{N}$ in terms of them:

$$\mathbf{S} = -\nabla\vartheta - \frac{\partial \mathbf{U}}{\partial t} = -c^2 \nabla\gamma - \frac{\partial(\gamma\mathbf{v})}{\partial t}, \qquad \mathbf{N} = \nabla\times\mathbf{U} = \nabla\times(\gamma\mathbf{v}). \tag{31}$$

Let us substitute (31) into the second equation in (30):

$$\nabla\times[\nabla\times(\gamma\mathbf{v})] = -\frac{\partial}{\partial t}\nabla\gamma - \frac{1}{c^2}\frac{\partial^2(\gamma\mathbf{v})}{\partial t^2} + \frac{4\pi\eta\gamma\rho_0\mathbf{v}}{c^2}. \tag{32}$$

The gauge condition of the 4-potential of the acceleration field (9) has the form: $\nabla^\mu u_\mu = 0$. In Minkowski space this relation is simplified:

$$\partial^\mu u_\mu = \frac{1}{c^2}\frac{\partial\vartheta}{\partial t} + \nabla\mathbf{U} = 0, \qquad \text{or} \qquad \frac{\partial\gamma}{\partial t} + \nabla\cdot(\gamma\mathbf{v}) = 0. \tag{33}$$

With regard to (33) we will transform the left side of (32):

$$\nabla\times[\nabla\times(\gamma\mathbf{v})] = \nabla(\nabla\cdot(\gamma\mathbf{v})) - \Delta(\gamma\mathbf{v}) = -\frac{\partial}{\partial t}\nabla\gamma - \Delta(\gamma\mathbf{v}).$$

Substituting this in (32), we obtain the wave equation:

$$\frac{1}{c^2}\frac{\partial^2(\gamma\mathbf{v})}{\partial t^2} - \Delta(\gamma\mathbf{v}) = \frac{4\pi\eta\gamma\rho_0\mathbf{v}}{c^2}. \tag{34}$$

According to (34) the velocity $\mathbf{v}$ of the fluid motion in the system must conform to the wave equation, that means that the velocity is given by the system's parameters and changes continuously in transition from one control volume to another.

The wave equation for the Lorentz factor follows from (31) and the first equation in (30) with regard to (33):

$$\frac{\partial^2 \gamma}{\partial t^2} - c^2 \Delta \gamma = 4\pi \eta \gamma \rho_0.$$

We can express the velocity **v** from (34) and substitute it in (29):

$$\frac{d}{dt}\left[\gamma \mathbf{v}\left(1 + \frac{p_0}{\rho_0 c^2} + \frac{\alpha}{c^2}\right)\right] = \mathbf{a}_m - \frac{1}{\gamma}\nabla\left(\frac{p_0}{\rho_0}\right) - \frac{\varsigma}{4\pi\eta\gamma^2 c^2 \rho_0^2}\left(\frac{\partial^2(\gamma \mathbf{v})}{\partial t^2} - c^2 \Delta(\gamma \mathbf{v})\right) + \frac{\omega}{\gamma \rho_0}\nabla(\nabla \cdot \mathbf{v}).$$

(35)

Let us find out the physical meaning of the last term in (35). The gauge condition (33) of the 4-potential of the acceleration field can be rewritten as follows:

$$\nabla \cdot \mathbf{v} = -\frac{1}{\gamma}\frac{d\gamma}{dt} = -\frac{1}{2c^2 \gamma^4}\frac{dv^2}{dt}.$$

Hence, provided $\gamma \approx 1$ we have:

$$\frac{\omega}{\gamma \rho_0}\nabla(\nabla \cdot \mathbf{v}) = -\frac{\omega}{2c^2 \gamma \rho_0}\nabla\left(\frac{1}{\gamma^4}\frac{dv^2}{dt}\right) \approx -\frac{\omega}{2c^2 \rho_0}\left(\frac{d}{dt}\nabla v^2 + (\nabla v^2 \cdot \nabla)\mathbf{v} + \nabla v^2 \times (\nabla \times \mathbf{v})\right). \quad (36)$$

The quantity $\frac{1}{2}\nabla v^2$ is the gradient of half of the squared velocity, that is the gradient of the kinetic energy per unit mass. This quantity is proportional to the acceleration, arising due to the dissipation of the kinetic energy of motion. The time derivative of $\nabla v^2$ leads to the rate of acceleration change. Other terms in (36) also have the dimension of the rate of acceleration change.

Thus in (35) viscosity is taken into account not only due to the motion velocity, but also due to the rate of acceleration change of the fluid motion.

**9.2. Comparison with the Navier-Stokes equation**

The vector Navier-Stokes equation in its classical form is usually used for non-relativistic description of the liquid motion and has the following form [6]:

$$\mathbf{a} = \frac{d\mathbf{v}}{dt} = \frac{\partial \mathbf{v}}{\partial t} + (\mathbf{v} \cdot \nabla)\mathbf{v} = \mathbf{a}_m - \frac{1}{\rho_0}\nabla p + \nu \Delta \mathbf{v} + \left(\frac{\xi}{\rho_0} + \frac{\nu}{3}\right)\nabla(\nabla \cdot \mathbf{v}), \qquad (37)$$

where $\mathbf{a}$ and $\mathbf{v}$ are the velocity and acceleration of an arbitrary point unit of liquid, $\rho_0$ is the mass density, $p$ is the pressure, $\nu$ is the kinematic viscosity coefficient, $\xi$ is the volume (bulk or second) viscosity coefficient, $\mathbf{a}_m$ is the acceleration produced by the mass forces in the liquid, and it is assumed that the coefficients $\nu$ and $\xi$ are constant in volume.

In (37) the velocity $\mathbf{v}$ depends not only on the time but also on the coordinates of the moving liquid unit. This allows us to expand the derivative $\frac{d\mathbf{v}}{dt}$ into the sum of two partial derivatives: time derivative $\frac{\partial \mathbf{v}}{\partial t}$ and space derivative $(\mathbf{v} \cdot \nabla)\mathbf{v}$, that is to apply the material (substantial) derivative.

Comparing (35) and (37) for the case of low velocities, when $\gamma$ tends to unity, and at sufficiently low pressure and viscosity, we obtain the kinematic viscosity coefficient:

$$\nu \approx \frac{\varsigma}{4\pi \eta \rho_0^2}. \qquad (38)$$

Since $\nu = \frac{\mu}{\rho_0}$, where $\mu$ is the dynamic viscosity coefficient, then we obtain:

$$\mu \approx \frac{\varsigma}{4\pi \eta \rho_0}.$$

In this ratio $\mu$ depends primarily on the fluid properties, and the coefficients $\varsigma$ and $\eta$ also depend on the parameters of the system under consideration. For example, if we study the liquid flow between two closely located plates, the coefficient $\varsigma$ is inversely proportional to the square of the distance between the plates.

The equality of the last terms in (35) and (37) implies:

$$\frac{\omega}{\rho_0} = \frac{\xi}{\rho_0} + \frac{\nu}{3} = \frac{\xi}{\rho_0} + \frac{\mu}{3\rho_0}, \qquad \omega = \xi + \frac{\mu}{3}. \tag{39}$$

The presence of $\xi$ and $\nu$ in (37) implies two causes of the rate of acceleration change: one of them is due to the fluid density variation because of the medium resistance and the other is due to the momentum variation of the fluid moving in a viscous medium.

### 9.3. The energy power

In Minkowski space the time component of the 4-velocity is equal to $u_0 = \gamma c$, the Christoffel symbols are zero, and (22) can be written as follows:

$$c^2 \frac{d\gamma}{dt} = \frac{1}{2\gamma} \frac{d(\gamma \mathbf{v})^2}{dt} = \mathbf{\Gamma} \cdot \mathbf{v} + \frac{\rho_{0q}}{\rho_0} \mathbf{E} \cdot \mathbf{v} + \mathbf{C} \cdot \mathbf{v} + \mathbf{X} \cdot \mathbf{v}. \tag{40}$$

We can also obtain (40), if in Minkowski space we multiply the equation of motion (23) by the velocity $\mathbf{v}$. We substitute in (40) the vector $\mathbf{C}$ from (25) and the vector $\mathbf{X}$ according to (A7) from Appendix A:

$$\frac{1}{2\gamma} \frac{d(\gamma \mathbf{v})^2}{dt} = \mathbf{\Gamma} \cdot \mathbf{v} + \frac{\rho_{0q}}{\rho_0} \mathbf{E} \cdot \mathbf{v} - \mathbf{v} \cdot \nabla \left( \frac{\gamma p_0}{\rho_0} \right) - \frac{\partial}{\partial t} \left( \frac{\gamma p_0 \mathbf{v}}{\rho_0 c^2} \right) \mathbf{v} - \mathbf{v} \cdot \nabla(\alpha \gamma) - \frac{1}{c^2} \frac{\partial(\alpha \gamma \mathbf{v})}{\partial t} \mathbf{v}.$$

The equivalent relation is obtained, if (29) is multiplied by the velocity $\mathbf{v}$:

$$\mathbf{v} \cdot \frac{d}{dt} \left[ \gamma \mathbf{v} \left( 1 + \frac{p_0}{\rho_0 c^2} + \frac{\alpha}{c^2} \right) \right] = \mathbf{v} \cdot \mathbf{a}_m - \frac{1}{\gamma} \mathbf{v} \cdot \nabla \left( \frac{p_0}{\rho_0} \right) - \frac{\varsigma}{\gamma \rho_0 c^2} \mathbf{v}^2 + \frac{\omega}{\gamma \rho_0} \mathbf{v} \cdot \nabla(\nabla \cdot \mathbf{v}). \tag{41}$$

The left side of (41) contains the rate of change of the kinetic energy per unit mass density (with contribution from the pressure and the dissipation function $\alpha$), and the right side contains the power of gravitational and electromagnetic forces $\mathbf{v} \cdot \mathbf{a}_m$ and the power of the pressure force. The term with the squared velocity in the right side of (41) is proportional to the kinetic energy, and the last term describes the power of the energy transformed during the fluid motion in a viscous medium with the effect of fluid compression and change of its density.

Another equivalent relation for the power of change in the energy of moving fluid is obtained by multiplying the velocity $\mathbf{v}$ by equation (35). In this case, in the right-hand side of (41) the Laplacian and the second partial time derivative appear. If we substitute the coefficients $\varsigma$ and $\omega$ with (38) and (39), we obtain the following:

$$\mathbf{v} \cdot \frac{d}{dt}\left[\gamma \mathbf{v}\left(1 + \frac{p_0}{\rho_0 c^2} + \frac{\alpha}{c^2}\right)\right] = \\ = \mathbf{v} \cdot \mathbf{a}_m - \frac{1}{\gamma} \mathbf{v} \cdot \nabla\left(\frac{p_0}{\rho_0}\right) - \frac{\nu}{\gamma^2 c^2} \mathbf{v} \cdot \left(\frac{\partial^2(\gamma \mathbf{v})}{\partial t^2} - c^2 \Delta(\gamma \mathbf{v})\right) + \frac{1}{\gamma \rho_0}\left(\xi + \frac{\mu}{3}\right) \mathbf{v} \cdot \nabla(\nabla \cdot \mathbf{v}).$$
(42)

### 9.4. Dependence of the velocity magnitude on the time

In this section we will make a conclusion about the nature of the kinetic energy change over time. For convenience, we will consider the co-moving reference frame in which the velocities $\mathbf{v}$ are relative velocities of motion of the fluid layers relative to each other. If we multiply all the terms in (27) by $\gamma$ and assume that the quantity $1 + \frac{p_0}{\rho_0 c^2} + \frac{\alpha}{c^2} \approx 1$, that is we neglect the contribution of the pressure energy density and the dissipation function as compared to the energy density at rest, we can write:

$$\gamma \frac{d(\gamma \mathbf{v})}{dt} = \gamma \mathbf{a}_m - \nabla\left(\frac{p_0}{\rho_0} + \alpha\right).$$

Assuming that $\nabla = \frac{d}{d\mathbf{r}}$, the velocity $\mathbf{v} = \frac{d\mathbf{r}}{dt}$, $\nabla(\nabla \cdot \mathbf{v})d\mathbf{r} = d(\nabla \cdot \mathbf{v})$, after multiplication by $d\mathbf{r}$ and integration, with regard to (28) for $\alpha$, we obtain:

$$\frac{1}{2}(\gamma \mathbf{v}_2)^2 - \frac{1}{2}(\gamma \mathbf{v}_1)^2 = \int \gamma \mathbf{a}_m \, d\mathbf{r} - \left[\left(\frac{p_0}{\rho_0}\right)_2 - \left(\frac{p_0}{\rho_0}\right)_1\right] - \frac{\varsigma}{c^2} \int \frac{1}{\rho_0} \mathbf{v} \, d\mathbf{r} + \omega \int \frac{1}{\rho_0} d(\nabla \cdot \mathbf{v}). \quad (43)$$

According to (43), the kinetic energy changes, when the work is carried out by the mass forces on the fluid, the fluid turns into a state with a different ratio $\dfrac{p_0}{\rho_0}$, in the fluid there is friction between the layers, and the velocity divergence is non-zero.

In (43) further simplification is possible, if we assume that in the process of integration the integrands change insignificantly and can be taken outside the integral sign. We will also use the continuity equation in the form:

$$\nabla \cdot \mathbf{v} = -\frac{1}{\gamma \rho_0} \frac{d}{dt}(\gamma \rho_0). \tag{44}$$

All this with regard to (38-39) gives:

$$\frac{1}{2}(\gamma \mathbf{v}_2)^2 - \frac{1}{2}(\gamma \mathbf{v}_1)^2 \approx \gamma \mathbf{a}_m \cdot \mathbf{r} - \left[\left(\frac{p_0}{\rho_0}\right)_2 - \left(\frac{p_0}{\rho_0}\right)_1\right] - \frac{4\pi \eta \mu}{c^2} \mathbf{v} \cdot \mathbf{r} - \frac{1}{\gamma \rho_0^2}\left(\xi + \frac{\mu}{3}\right)\frac{d}{dt}(\gamma \rho_0). \tag{45}$$

The left side of (45) contains the change of kinetic energy per unit mass, which occurs due to the velocity change from $\mathbf{v}_1$ to $\mathbf{v}_2$. The kinetic energy increases if in the right side the projection $\mathbf{a}_m$ of the mass forces' acceleration on the displacement vector $\mathbf{r}$ has a positive sign. Meanwhile the second, third and fourth terms in the right side have a negative sign. This means that the motion energy dissipation is proportional to the increase in pressure during the fluid motion, the velocity and the motion distance, as well as to the increase in the fluid density that prevents from free motion.

The scalar product $\mathbf{v} \cdot \mathbf{r} = v^2 t$, where $t$ denotes the time of motion of one layer relative to another, does not vanish during the curvilinear or rotational motion of the layers of fluid or liquid. Therefore in the moving fluid vortices and turbulence can easily occur. This is contributed by the fact that the terms in the right side of (45) can influence each other. For example, in the areas of high pressure the fluid streamlines bent, and the temperature changes the density in the last term in (45). Turbulence can be characterized as a method of transferring the energy of linear motion of the fluid into the rotary motion of different scales.

Equation (45) can be rewritten so as to move all the terms, depending on the velocity, to the left side. Assuming $\mathbf{a}_m \cdot \mathbf{r} = \mathbf{a}_m \cdot \mathbf{v}\, t = a_m \cdot v\, t \cos \beta$, where $\beta$ is the angle between the velocity and the acceleration $\mathbf{a}_m$, we find a quadratic equation for the velocity as a function of the time $t$ and other parameters:

$$\frac{1}{2}\gamma^2 v^2 + \frac{4\pi\eta\mu}{c^2}v^2 t - \gamma\, \mathbf{a}_m \cdot \mathbf{v}\, t\cos\beta = -\frac{p_0}{\rho_0} - \frac{1}{\gamma\rho_0^2}\left(\xi + \frac{\mu}{3}\right)\frac{d}{dt}(\gamma\rho_0) + \text{const}.$$

The constant in the right side specifies the initial condition of motion. The solution of this equation allows us to estimate the change of the velocity magnitude over time.

## 10. Conclusion

For the case of constant coefficients of viscosity we showed that the Navier-Stokes equation of motion of the viscous compressible liquid can be derived using the 4-potential of the energy dissipation field, dissipation tensor and dissipation stress-energy tensor. First we wrote the equations of motion (15) in a general form, then expressed them in (22-23) through the strengths of the gravitational and electromagnetic fields, the strengths of the pressure field and energy dissipation field. The 4-potential of the dissipation field includes the dissipation function $\alpha$ and the associated scalar potential $\varepsilon$ of the dissipation field. The quantity $\alpha$ can be selected so that in the equation (27) for the fluid acceleration the dependence on the velocity of the fluid motion appears, associated with viscosity, when deceleration of the fluid is proportional to the relative velocity of its motion. We can also take into account the dependence on the rate of acceleration change over time. This gives us the equation (29).

Then we analyzed the wave equation for the velocity field (34) and expressed the velocity from to substitute it in (29). The resulting equation (35) coincides almost exactly with the Navier-Stokes equation (37). One difference is that in the acceleration from pressure the mass density $\rho_0$ in the expression $\nabla\left(\dfrac{p_0}{\rho_0}\right)$ is under the gradient sign, and in (37) $\rho_0$ is taken outside the gradient sign. The second difference is due to the fact that in (35) there is an additional term in the form $\dfrac{\partial^2(\gamma\mathbf{v})}{\partial t^2}$. This term is proportional to the rate of acceleration change over time and describes the phenomena, in which the change of the medium properties affecting viscosity occurs in a specified time frame.

In addition, in (35) we took into account the relativistic corrections of the Lorentz factor $\gamma$, as well as the fluid acceleration dependence on the acceleration of the mass-energy of the pressure field and dissipation field ( in square brackets in the left side of ( 35) ).

The directed kinetic energy of motion of the fluid in a viscous medium can dissipate into the random motion of the particles of the surrounding medium and be converted into heat. The inverse process is also possible, when heating of the medium leads to a change in the state of the fluid motion. In section 9.3. we introduced the differential equations of the change in the system's kinetic energy and its conversion into other energy forms, including the dissipation field energy. These equations are not completely independent, since they are obtained by scalar multiplication of the equation of motion by the fluid velocity $\mathbf{v}$.

The dissipation stress-energy tensor $Q^{\alpha\beta}$ is represented in (16) and its invariants are represented in (A8) in Appendix A. In section 9.1. the dissipation function is given by formula (28): $\alpha = \dfrac{\varsigma}{c^2}\int \dfrac{1}{\rho_0}\mathbf{v}\,d\mathbf{r} - \omega\int \dfrac{1}{\rho_0}\nabla(\nabla\cdot\mathbf{v})\,d\mathbf{r}$. This function depends on the distance traveled by the fluid relative to the surrounding moving medium, and can be considered as a function of the time of motion with respect to the reference frame, which is at the average co-moving with the fluid in this small control volume of the system. With the help of the known quantity $\alpha$ we can calculate according the formulas (A7) the vectors $\mathbf{X}$ and $\mathbf{Y}$ and therefore determine the components of the tensor $Q^{\alpha\beta}$. In particular, the volume integral of the component $Q^{00}$ of this tensor allows us to consider all the energy that is transferred by the moving fluid to the surrounding medium, in the form of dissipation field energy, and the components $Q^{0i}$ define the vector $\mathbf{Z} = cQ^{0i}$ as the energy flux density of the dissipation field.

Under the assumptions made the Navier-Stokes equation (37) reduces to equation (27), wherein the acceleration depends, besides the mass forces, on the sum of two gradients – the dissipation function $\alpha$ and the quantity $\dfrac{p_0}{\rho_0}$. Equation (27) has such a form that this equation should have smooth solutions, if there are no discontinuities in the pressure or the dissipation function $\alpha$. If we consider condition (28) and formula (36) as valid, the gradient of $\alpha$ will also be a smooth function.

Instead of moving from equation (27) to equation (35), which is similar to the Navier-Stokes equation (37), we can act in another way. Differential equation (27) is an equation to determine the velocity field $\mathbf{v}$. In this equation, there are at least three more unknown functions: the pressure field $p_0$, the mass density $\rho_0$, and the dissipation function $\alpha$. Therefore, it is necessary to add to (27) at least three equations in order to close the system of equations and make it solvable in principle. One of such equations is the continuity equation (12) in the form of (44), which relates the density and velocity. In order to determine the dissipation function $\alpha$ we have introduced the wave equation (A10) in Appendix A. The pressure distribution in the system can be found from the wave equation (B4) in Appendix B.

In equation (27) there is also acceleration $\mathbf{a}_m$, arising due to the action of mass forces. This acceleration depends on the gravitational field strength $\mathbf{\Gamma}$, torsion field $\mathbf{\Omega}$, electric field strength $\mathbf{E}$, magnetic field $\mathbf{B}$ and charge density $\rho_{0q}$:

$$\mathbf{a}_m = \mathbf{\Gamma} + [\mathbf{v} \times \mathbf{\Omega}] + \frac{\rho_{0q}}{\rho_0}\mathbf{E} + \frac{\rho_{0q}}{\rho_0}[\mathbf{v} \times \mathbf{B}].$$

For each of these quantities there are special equations used to define them. For example, the gravitational field equations (the Heaviside equations) can be represented according to [13] as follows:

$$\nabla \cdot \mathbf{\Gamma} = -4\pi G \gamma \rho_0, \quad \nabla \times \mathbf{\Omega} = \frac{1}{c^2}\frac{\partial \mathbf{\Gamma}}{\partial t} - \frac{4\pi G \gamma \rho_0 \mathbf{v}}{c^2}, \quad \nabla \cdot \mathbf{\Omega} = 0, \quad \nabla \times \mathbf{\Gamma} = -\frac{\partial \mathbf{\Omega}}{\partial t}. \quad (46)$$

Equations (46) are derived in [11] from the principle of least action and are similar in their form to Maxwell equations, which are used to calculate $\mathbf{E}$ and $\mathbf{B}$. Finally, the charge density $\rho_{0q}$ can be related to the velocity by means of the equation of the electric charge continuity:

$$\nabla \cdot \mathbf{v} = -\frac{1}{\gamma \rho_{0q}}\frac{d}{dt}(\gamma \rho_{0q}). \quad (47)$$

Thus, the set of equations (27), (44), (A10), (B4), (46), (47) together with Maxwell equations is a complete set, which is sufficient to solve the problem of motion of viscous compressible and charged fluid in the gravitational and electromagnetic fields.

**Appendix A. Properties of the dissipation field**

The components of the antisymmetric tensor of the dissipation field are obtained from relation (3) using the relation (2). We will introduce the following notations:

$$h_{0i} = \partial_0 \lambda_i - \partial_i \lambda_0 = \frac{1}{c} X_i, \qquad h_{ij} = \partial_i \lambda_j - \partial_j \lambda_i = -Y_k, \qquad (A1)$$

where the indices $i, j, k$ form triplets of non-recurrent numbers of the form 1,2,3 or 3,1,2 or 2,3,1; the 3-vectors **X** and **Y** can be written by components: $\mathbf{X} = X_i = (X_1, X_2, X_3) = (X_x, X_y, X_z)$; $\mathbf{Y} = Y_i = (Y_1, Y_2, Y_3) = (Y_x, Y_y, Y_z)$.

Using these notations the tensor $h_{\mu\nu}$ can be represented as follows:

$$h_{\mu\nu} = \begin{pmatrix} 0 & \frac{X_x}{c} & \frac{X_y}{c} & \frac{X_z}{c} \\ -\frac{X_x}{c} & 0 & -Y_z & Y_y \\ -\frac{X_y}{c} & Y_z & 0 & -Y_x \\ -\frac{X_z}{c} & -Y_y & Y_x & 0 \end{pmatrix}. \qquad (A2)$$

The same tensor with contravariant indices equals: $h^{\alpha\beta} = g^{\alpha\mu} g^{\nu\beta} h_{\mu\nu}$. In Minkowski space the metric tensor does not depend on the coordinates, in which case for the dissipation tensor it follows:

$$h^{\alpha\beta} = \eta^{\alpha\mu} \eta^{\nu\beta} h_{\mu\nu} = \begin{pmatrix} 0 & -\frac{X_x}{c} & -\frac{X_y}{c} & -\frac{X_z}{c} \\ \frac{X_x}{c} & 0 & -Y_z & Y_y \\ \frac{X_y}{c} & Y_z & 0 & -Y_x \\ \frac{X_z}{c} & -Y_y & Y_x & 0 \end{pmatrix}. \qquad (A3)$$

We can express the dissipation field equations (8) in Minkowski space in terms of the vectors $\mathbf{X}$ and $\mathbf{Y}$ using the 4-vector of mass current: $J^\mu = \rho_0 u^\mu = \rho_0(\gamma c, \gamma \mathbf{v})$, where $\gamma = \dfrac{1}{\sqrt{1-v^2/c^2}}$. Replacing in (8) the covariant derivatives $\nabla_\beta$ with the partial derivatives $\partial_\beta$ we find:

$$\nabla \cdot \mathbf{X} = 4\pi\tau\gamma\rho_0, \quad \nabla \times \mathbf{Y} = \frac{1}{c^2}\frac{\partial \mathbf{X}}{\partial t} + \frac{4\pi\tau\gamma\rho_0 \mathbf{v}}{c^2}, \quad \nabla \cdot \mathbf{Y} = 0, \quad \nabla \times \mathbf{X} = -\frac{\partial \mathbf{Y}}{\partial t}. \qquad (A4)$$

If we multiply scalarly the second equation in (A4) by $\mathbf{X}$ and the fourth equation — by $-\mathbf{Y}$ and then sum up the results, we will obtain the following:

$$-\nabla \cdot [\mathbf{X} \times \mathbf{Y}] = \frac{1}{2c^2}\frac{\partial(X^2 + c^2 Y^2)}{\partial t} + \frac{4\pi\tau\gamma\rho_0 \mathbf{v} \cdot \mathbf{X}}{c^2}. \qquad (A5)$$

Equation (A5) comprises the Poynting theorem applied to the dissipation field. The meaning of this differential equation that if dissipation of the energy of moving fluid particles takes place in the system, then the divergence of the field dissipation flux is associated with the change of the dissipation field energy over time and the power of the dissipation energy density. Relation (A5) in a covariant form is written as the time component of equation (17):

$$\nabla_\beta Q^{0\beta} = -h^{0\beta} J_\beta.$$

If we substitute (A2) into (17), we can express the scalar and vector components of the 4-force density of the dissipation field:

$$(f_0)_d = h_{0\sigma} J^\sigma = \frac{\gamma\rho_0}{c}\mathbf{X} \cdot \mathbf{v}, \qquad (f_i)_d = h_{i\sigma} J^\sigma = -\gamma\rho_0\left(\mathbf{X} + [\mathbf{v} \times \mathbf{Y}]\right). \qquad (A6)$$

The vector $\mathbf{X}$ has the dimension of an ordinary 3-acceleration, and the dimension of the vector $\mathbf{Y}$ is the same as that of the frequency.

Substituting the 4-potential of the dissipation field (2) in the definition (A1), in Minkowski space we find:

$$\mathbf{X} = -\nabla \varepsilon - \frac{\partial \mathbf{\Theta}}{\partial t} = -\nabla(\alpha\gamma) - \frac{1}{c^2}\frac{\partial(\alpha\gamma\mathbf{v})}{\partial t}, \qquad \mathbf{Y} = \nabla \times \mathbf{\Theta} = \frac{1}{c^2}\nabla \times (\alpha\gamma\mathbf{v}). \qquad (A7)$$

The vector $\mathbf{X}$ is the dissipation field strength, and the vector $\mathbf{Y}$ is the solenoidal vector of the dissipation field. Both vectors depend on the dissipation function $\alpha$, which in turn depends on the coordinates and time. In real fluids there is always internal friction, $\alpha \neq 0$, and the vectors $\mathbf{X}$ and $\mathbf{Y}$ are also not equal to zero.

We can substitute the tensors (A2) and (A3) in (16) and express the stress-energy tensor of the dissipation field $Q^{\alpha\beta}$ in terms of the vectors $\mathbf{X}$ and $\mathbf{Y}$. We will write here the expression for the tensor invariant $h_{\mu\nu}h^{\mu\nu}$ and for the time components of the tensor $Q^{\alpha\beta}$:

$$h_{\mu\nu}h^{\mu\nu} = -\frac{2}{c^2}(X^2 - c^2 Y^2), \qquad Q^{00} = \frac{1}{8\pi\tau}(X^2 + c^2 Y^2), \qquad Q^{0i} = \frac{c}{4\pi\tau}[\mathbf{X}\times\mathbf{Y}]. \qquad (A8)$$

The component $Q^{00}$ defines the energy density of the dissipation field in the given volume, and the vector $\mathbf{Z} = cQ^{0i} = \frac{c^2}{4\pi\tau}[\mathbf{X}\times\mathbf{Y}]$ defines the energy flux density of the dissipation field.

If we substitute $\mathbf{X}$ from (A7) into the first equation in (A4), and take into account the gauge of the 4-potential (9) as follows:

$$\partial^\mu \lambda_\mu = \frac{1}{c^2}\frac{\partial \varepsilon}{\partial t} + \nabla \cdot \mathbf{\Theta} = 0, \quad \text{or} \quad \frac{\partial(\alpha\gamma)}{\partial t} + \nabla \cdot (\alpha\gamma\mathbf{v}) = 0, \qquad (A9)$$

we will obtain the wave equation for the scalar potential:

$$\frac{1}{c^2}\frac{\partial^2 \varepsilon}{\partial t^2} - \Delta\varepsilon = 4\pi\tau\gamma\rho_0, \quad \text{or} \quad \frac{1}{c^2}\frac{\partial^2(\alpha\gamma)}{\partial t^2} - \Delta(\alpha\gamma) = 4\pi\tau\gamma\rho_0. \qquad (A10)$$

From (A7), (A9) and the second equation in (A4) the wave equation follows for the vector potential of the dissipation field:

$$\frac{1}{c^2}\frac{\partial^2 \mathbf{\Theta}}{\partial t^2} - \Delta \mathbf{\Theta} = \frac{4\pi\tau\gamma\rho_0 \mathbf{v}}{c^2}, \quad \text{or} \quad \frac{1}{c^2}\frac{\partial^2 (\alpha\gamma\mathbf{v})}{\partial t^2} - \Delta(\alpha\gamma\mathbf{v}) = 4\pi\tau\gamma\rho_0 \mathbf{v}.$$

### Appendix B. Pressure field equations

Four vector equations for the pressure field components within the special theory of relativity were presented in [7] as the consequence of the action function variation:

$$\nabla \cdot \mathbf{C} = 4\pi\sigma\gamma\rho_0, \quad \nabla \times \mathbf{I} = \frac{1}{c^2}\frac{\partial \mathbf{C}}{\partial t} + \frac{4\pi\sigma\gamma\rho_0 \mathbf{v}}{c^2}, \quad \nabla \cdot \mathbf{I} = 0, \quad \nabla \times \mathbf{C} = -\frac{\partial \mathbf{I}}{\partial t}. \quad (B1)$$

The vector of the pressure field strength $\mathbf{C}$ and the solenoidal vector $\mathbf{I}$ are determined with the 4-potential of the pressure field $\pi_\mu = \frac{p_0}{\rho_0 c^2} u_\mu = \left(\frac{\wp}{c}, -\mathbf{\Pi}\right)$ according to the formulas:

$$\mathbf{C} = -\nabla\wp - \frac{\partial \mathbf{\Pi}}{\partial t} = -\nabla\left(\frac{\gamma p_0}{\rho_0}\right) - \frac{\partial}{\partial t}\left(\frac{\gamma p_0 \mathbf{v}}{\rho_0 c^2}\right), \quad (B2)$$

$$\mathbf{I} = \nabla \times \mathbf{\Pi} = \nabla \times \left(\frac{\gamma p_0 \mathbf{v}}{\rho_0 c^2}\right).$$

The 4-potential gauge according to (9) in the form $\nabla^\mu \pi_\mu = 0$ in Minkowski space is transformed into the expression $\partial^\mu \pi_\mu = 0$. Substituting here the expression for the 4-potential of the pressure field, we obtain:

$$\partial^\mu \pi_\mu = \frac{1}{c^2}\frac{\partial \wp}{\partial t} + \nabla \cdot \mathbf{\Pi} = 0, \quad \text{or} \quad \frac{\partial}{\partial t}\left(\frac{\gamma p_0}{\rho_0}\right) + \nabla \cdot \left(\frac{\gamma p_0 \mathbf{v}}{\rho_0}\right) = 0. \quad (B3)$$

Substituting (B2) into the first equation in (B1) and using (B3), we obtain the wave equation for calculation of the scalar potential of the pressure field:

$$\frac{1}{c^2}\frac{\partial^2 \wp}{\partial t^2} - \Delta\wp = 4\pi\sigma\gamma\rho_0, \quad \text{or} \quad \frac{1}{c^2}\frac{\partial^2}{\partial t^2}\left(\frac{\gamma p_0}{\rho_0}\right) - \Delta\left(\frac{\gamma p_0}{\rho_0}\right) = 4\pi\sigma\gamma\rho_0. \quad (B4)$$

The wave equation for the vector potential of the pressure field follows from (B2), (B3) and the second equation in (B1):

$$\frac{\partial^2 \mathbf{\Pi}}{\partial t^2} - c^2 \Delta \mathbf{\Pi} = 4\pi \sigma \gamma \rho_0 \mathbf{v}, \quad \text{or} \quad \frac{1}{c^2} \frac{\partial^2}{\partial t^2}\left(\frac{\gamma p_0 \mathbf{v}}{\rho_0}\right) - \Delta\left(\frac{\gamma p_0 \mathbf{v}}{\rho_0}\right) = 4\pi \sigma \gamma \rho_0 \mathbf{v}.$$